\documentclass{article}

    \usepackage[preprint]{neurips_2023}

\usepackage[utf8]{inputenc} 
\usepackage[T1]{fontenc}    
\usepackage{hyperref}       
\usepackage{url}            
\usepackage{booktabs}       
\usepackage{amsfonts}       
\usepackage{nicefrac}       
\usepackage{microtype}      
\usepackage{xcolor}         
\usepackage{multirow}
\usepackage{multicol}
\PassOptionsToPackage{numbers, compress}{natbib}

\usepackage{graphicx}
\usepackage{amsmath}
\usepackage{amssymb}
\usepackage{booktabs}
\usepackage{color}
\usepackage{pythonhighlight}

\title{Rethinking Learned Image Compression: Context is All You Need }

\author{%
  Jixiang Luo \thanks{Work done at Sensetime Research} \\
  \texttt{jixiangluo85@gmail.com} \\
}

\begin{document}

\maketitle

\begin{abstract}
  Since LIC has made rapid progress recently compared to traditional methods, this paper attempts to discuss the question about 'Where is the boundary of Learned Image Compression(LIC)?'. Thus this paper splits the above problem into two sub-problems:1$)$Where is the boundary of rate-distortion performance of PSNR? 2$)$ How to further improve the compression gain and achieve the boundary? Therefore this paper analyzes the effectiveness of scaling parameters for encoder, decoder and context model, which are the three components of LIC. Then we conclude that scaling for LIC is to scale for context model and decoder within LIC. Extensive experiments demonstrate that overfitting can actually serve as an effective context. By optimizing the context, this paper further improves PSNR and achieves state-of-the-art performance, showing a performance gain of $14.39\%$ with BD-RATE over VVC.
\end{abstract}

\section{Introduction}
The exponential growth of digital imagery in recent years has driven the need for efficient storage and transmission solutions, while traditional methods like JPEG~\cite{jpeg}, BPG~\cite{bpg}, WebP~\cite{webp} can not meet the demands. With the proliferation of high-resolution cameras and the widespread adoption of multimedia applications, the size of image data has surged, presenting significant challenges in terms of storage capacity and bandwidth requirements. Traditional image compression techniques, such as JPEGXL~\cite{alakuijala2019jpeg} including lossy and lossless JPEG recompression,  AVIF~\cite{barman2020evaluation} extracted from AV1~\cite{av1}, HEIF~\cite{lainema2016hevc} extracted from HEVC~\cite{hevc}, have been the cornerstone of addressing these challenges. However, these methods often struggle to achieve the desired balance between compression efficiency and preservation of image quality. While these approaches have theoretical and mathematical explanations, they are all based on handcrafted designs of individual modules, lacking joint optimization of objectives to achieve an optimal Pareto frontier. Additionally, they fail to meet the increasing demands for higher image quality, such as lower bit rates and higher objective and subjective image quality. 

In response to these challenges, deep learning-based image compression techniques have emerged as a promising alternative. By harnessing the power of neural networks, these methods offer several advantages over traditional compression algorithms.  Deep learning-based approaches, which encompass \textbf{encoder, decoder, quantization and context model }, have demonstrated remarkable progress in recent years, significantly advancing the state-of-the-art in image compression. This paper comprehensively reviews the progress of LIC, provides intuitive theoretical explanations, and explores the boundaries of LIC objective quality. One of the key advancements in deep learning-based image compression is the incorporation of encoder and decoder architectures. The encoder component compresses the input image into a latent representation, while the decoder component reconstructs the original image from this representation. Additionally, context models are employed to capture high-level semantic information, ensuring that important features are preserved during the compression process.
The backbone of LIC, comprising the encoder and decoder, primarily consists of several distinct architectural components, including convolutional neural networks (CNNs)~\cite{imagecnn6, imagecnn7, agustsson2019generative, he2021checkerboard, luo2020noise, li2020spatial}, recurrent neural networks (RNNs)~\cite{toderici2017full, islam2021image, toderici2015variable, lin2020spatial}, transformers~\cite{wang2022end, lu2021transformer, liu2023learned, kao2023transformer, jeny2022efficient}, and frequency decomposition methods~\cite{li2023frequency, choi2022frequency, zhang2024end, gao2021neural, liu2021image}. Despite the variety of structures available, the current trend indicates that higher complexity leads to greater compression gains. Correspondingly, most solutions tend to be associated with transformers to some extent. However, combining CNNs with transformers remains the optimal approach~\cite{li2023frequency, liu2023learned}. However, the difference lies in the usage: the former employs transformers in the context, while the latter integrates them into the backbone. Both approaches can enhance performance, but they also introduce varying degrees of computational complexity. Due to the high complexity of transformers in the backbone, some papers focus on optimizing the context module specifically~\cite{li2020spatial, li2020efficient,yuan2021learned,he2021checkerboard,minnen2020channel,elic}. This optimization can significantly enhance the performance of the entire LIC community with great efficiency. The most crucial aspect involves modeling the relationships between latents with spatial and channel dimensions . Currently, the prevailing approach is to simulate the distribution of latents based on conditional probability, thereby obtaining the conditional entropy of latents. The boundary of conditional entropy is closely related to the selection of conditions. Hence, we cannot accurately estimate the performance boundary of LIC. Given two variables $\mathcal{X}, \mathcal{Y}$, we have $H(\mathcal{Y}|\mathcal{X})<H(\mathcal{Y})$, where $H(*)$ is the entropy of variables. $H(\mathcal{Y}|\mathcal{X}) \sim \frac{H(\mathcal{Y})}{\rho(\mathcal{X}, \mathcal{Y})} $, The conditional entropy is related to the correlation $\rho$ between two variables. If the correlation between two variables is higher, the conditional entropy will be smaller, and vice versa.

\begin{table}
\centering
\caption{The rate-distortion performance of DCVC-FM~\cite{li2024neural} and ELIC~\cite{elic} on HEVC-B sequences~\cite{bossen2013common} with PSNR in RGB colorspace. I and P are the frame types defined by DCVC-FM, while ELIC takes all frames as image or I frames}
\label{tab:video}
\scalebox{0.95}{
\begin{tabular}{c|c|c|c|c|c|c|c}
\hline \hline
methods & type & & BQTerrace & BasketballDrive & Cactus & Kimono & ParkScene \\ \hline \hline
\multirow{4}{*}{DCVC-FM}& \multirow{2}{*}{I}& bpp & 0.0362 & 0.0156 & 0.0255 & 0.0157 & 0.0165 \\ \cline{3-8}
& & PSNR & 28.7238 & 33.6617 & 30.0544 & 33.0646 & 30.2771 \\ \cline{3-8} \cline{2-8}
& \multirow{2}{*}{P} & bpp & 0.0007 & 0.0015 & 0.0011 & 0.0011 & 0.0008 \\ \cline{3-8}
& & PSNR & 29.2897 & 29.3356 & 28.8999 & 30.5414 & 28.8406 \\ \cline{3-8} \hline \hline
\multirow{4}{*}{ELIC}& \multirow{2}{*}{I}& bpp & 0.0201 & 0.0096 & 0.0136 & 0.0090 & 0.0085 \\ \cline{3-8}
& & PSNR & 25.7859 & 29.7820 & 26.8106 & 29.9608 & 27.6546 \\ \cline{2-8}
&  \multirow{2}{*}{P}& bpp & 0.0193 & 0.0109 & 0.0133 & 0.0089 & 0.0089 \\ \cline{3-8}
& & PSNR & 26.1313 & 29.1154 & 26.7629 & 30.1376 & 27.2406 \\ \hline  \hline
\multirow{4}{*}{ELIC}& \multirow{2}{*}{I}& bpp & 0.1757 & 0.0487 & 0.1043 & 0.0618 & 0.1018 \\ \cline{3-8}
& & PSNR & 32.1260 & 33.4375 & 30.4493 & 34.0548 & 30.0388 \\ \cline{2-8}
&  \multirow{2}{*}{P}& bpp & 0.1609 & 0.0592 & 0.1047 & 0.0590 & 0.1100 \\ \cline{3-8}
& & PSNR & 32.4425 & 32.6736 & 30.4104 & 34.1136 & 30.0649 \\ \hline \hline
\end{tabular}}
\end{table}

Therefore, we define pixel level, structure level, and semantic level to discuss the compression limits of LIC. At the structure level, the overall contours or edges of the image are maintained consistently with the ground truth, while other areas can vary, thus improving the compression rate. A deeper level of semantic abstraction involves an overall understanding of the image, enabling unimaginable compression levels. For example, consider compressing an image with a clean background and a running Shiba Inu. The pixel level focuses on every pixel, the structure level focuses on the outline of the dog, but the semantic level abstracts it into a single sentence. Since this sentence requires only a few bytes of storage, it achieves extreme semantic compression. Since the structure level and semantic level can be attributed to fields such as image generation~\cite{rombach2022high, dhariwal2021diffusion} and image understanding~\cite{dosovitskiy2020image,betker2023improving}, and supervised generation faces other issues such as fidelity, which are beyond the scope of this discussion, the limits discussed in this paper are confined to the pixel level only.

In addition to the encoder and decoder, quantization is also a significant consideration. On one hand, entropy coders require a fixed finite symbol set, and on the other hand, quantization is an effective means of lossy compression to increase gains. This mainly involves scalar quantization~\cite{ cai2018deep, zhong2020channel, sun2022improving} and vector quantization~\cite{guo2021soft,kudo2022lvq, zhang2023lvqac}. 
After quantization, entropy coding is performed. Since entropy coding is a non-differentiable process, estimating the mean and variance during training and then using a Gaussian distribution to estimate the training bitrate is necessary. This is because the Gaussian distribution is the maximum entropy distribution~\cite{park2009maximum, lisman1972note}  constrained by the mean and variance, and it can be computed conveniently and quickly. Under the same constraints, the Laplace distribution also constrains the mean and variance, but its variance must be even. Its constraint property is slightly stricter than that of the Gaussian distribution. However, in some experiments, the Laplace distribution does not perform better than the Gaussian distribution even Logistic distribution in Fig.4 of ~\cite{imagecnn7}.
\begin{figure}[t]
    \centering
\includegraphics[width=0.9\linewidth]{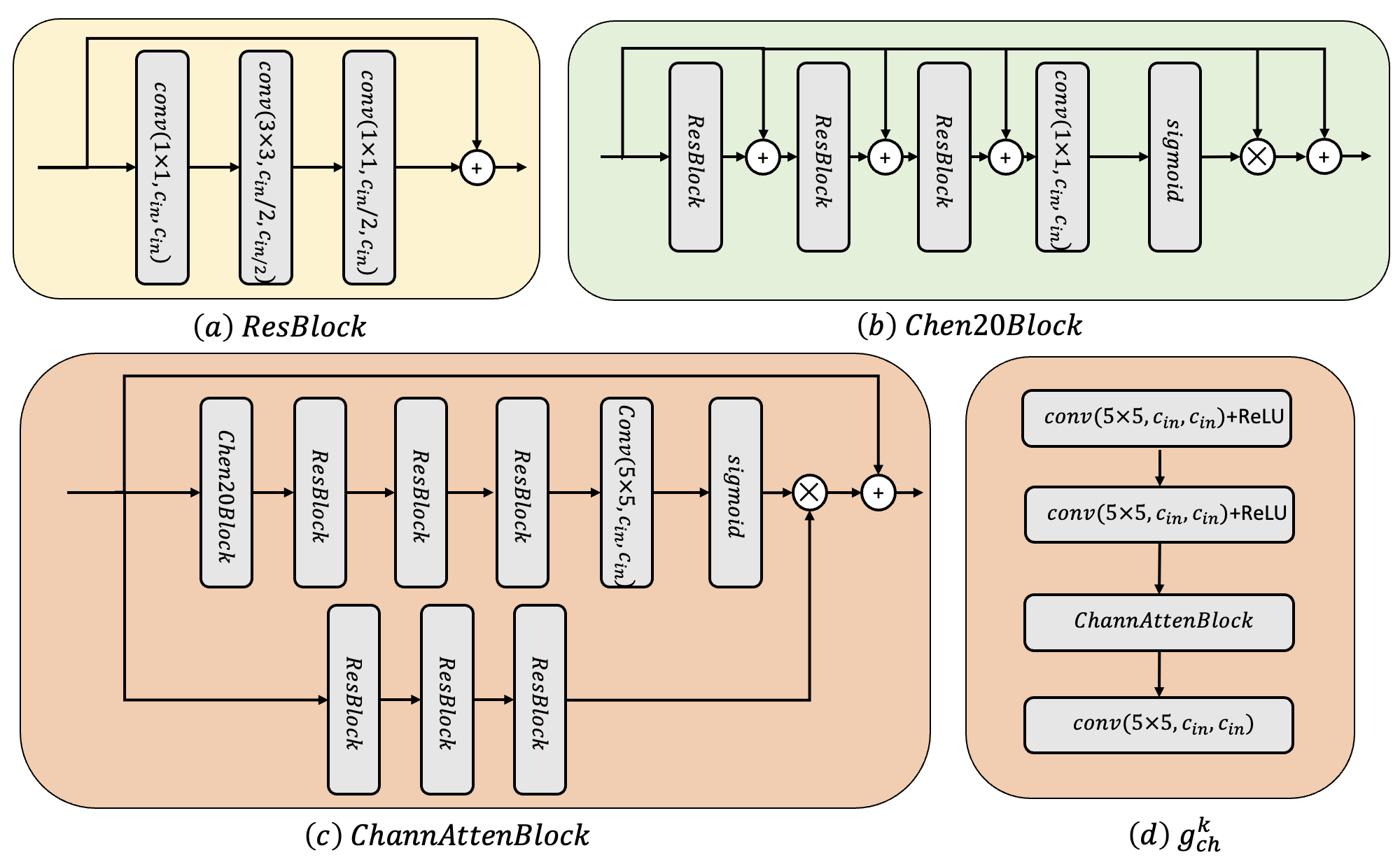}
    \caption{Channel attention for context model. $1\times1,  5\times5$ are the kernel size of convolutional layer, ReLU and softmax are the activation function. $c_{in}$ in the number of input channel}
    \label{fig:channattenblock}
\end{figure}

By further extending the dimensions of LIC, we can obtain a broader Pareto optimization frontier, encompassing \textbf{rate-distortion-perceptuality-performance-practicality(RDPPP)}, where perceptuality means subjective quality, performance includes some downstreaming task matrices(accuracy, precision, recall), and practicality includes model size, memory and GPU occupancy, runtime speed, codec compatibility, among other factors. Since the boundary of PSNR cannot be accurately explored, another important direction is the optimization of the subjective quality~\cite{he2022po, luo2024super, mentzer2020high, patel2021saliency, agustsson2023multi} and integration with downstream tasks~\cite{chamain2021end, choi2018deep, duan2020video, hu2020towards, kawawa2022recognition, xiao2022identity, luo2024compressible, feng2023semantically, le2021image, le2021learned, sun2020semantic} in LIC.
Researchers have been exploring the Pareto frontier of related studies from various dimensions to approach the limits of LIC. ~\cite{blau2019rethinking} exploits the Rate-Distortion-Perception, and ~\cite{yu2022evaluating} considers the practicality with regard to speed and hardware utility. Yet, from an information theory perspective, determining an upper bound for the conditional entropy of images remains elusive due to their inherent smoothness and low-rank nature. This complexity underscores the intricate nature of the problem and highlights the need for comprehensive approaches to push the performance boundary of LIC. 
However, due to the complexity of models and computational costs, it has been challenging to deploy LIC at both the edge and cloud devices. Therefore, while ensuring the performance of LIC, model simplification is also an important direction to consider. ~\cite{liao2022efficient} deployed  efficient decoder to improve decoder spped. ~\cite{yu2022evaluating, van2024mobilenvc} utilizes much more engineering optimization methods. ~\cite{fang2023fully, sun2022q, sun2020end, shi2023rate} quantized full network to reduce calculations. 
To further simplify the model, LIC is evolved from a single-rate model to a variable-rate model, allowing one model to cover low, medium, and high-rate points~\cite{toderici2015variable, choi2019variable, kao2023transformer, chen2020variable, song2021variable, sun2021interpolation, jiang2022online, lee2022selective}. 

 However, subjective quality, performance with  downstreaming task and practicality, being a challenging aspect to quantify, are beyond the scope of this paper, which only emphasize encoder, decoder, quantization and context model. This paper seeks to address the objective metric boundary (PSNR) of LIC by integrating the analysis of scaling laws~\cite{hoffmann2022training, caballero2022broken, henighan2020scaling} to explore their impact on LIC. The scaling law includes number of parameters, dataset size, computing cost, loss. 
 While increasing the training dataset and model parameters does not necessarily produce significant improvements in LIC. Therefore, this paper explores the performance of LIC by expanding the number of channels to increase model capacity, with a particular focus on examining the effects of overfitting in the context of dataset expansion.


\begin{figure}[t]
    \centering
    \includegraphics[width=0.9\linewidth]{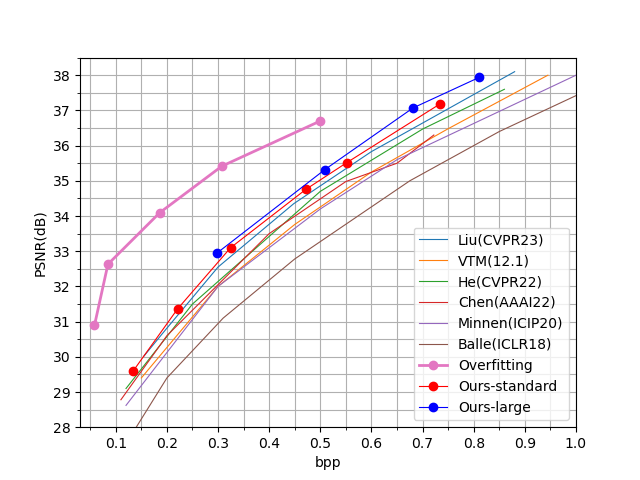}
    \caption{PSNR of Kodak dataset."Overfitting" means models trained at Kodak dataset.}
    \label{fig:kodak_psnr}
\end{figure}

\begin{table}[ht]
\caption{BD-RATE and BD-PSNR ~\cite{bd-rate} over VTM-12.1 at Kodak dataset}
\scalebox{0.9}{
\begin{tabular}{c|c|c|c|c|c|c}
\hline
    \%  & Ours-large & Ours   & Liu(CVPR23) & He(CVPR22) & Chen(AAAI22) & Minnen(ICIP20) \\ \hline \hline
BD-RATE & 16.91 & -14.39 & -11.06      & -6.87      & -4.09        & 1.43           \\ \hline
BD-PSNR & 93.12 & 69.22  & 54.12       & 31.02      & 17.96        & -6.36          \\ \hline \hline
\end{tabular}}
\label{bd-rate}
\end{table}

\section{Problem Formatting}
ELIC~\cite{elic} utilizes a convolutional-based encoder and decoder, along with both intra-channel and inter-channel context modeling. Specifically, the channels are unevenly divided into five groups, where the earlier groups serve as the context for the later ones. Within each group, a checkerboard pattern is employed so that one subset of the checkerboard serves as the context for the other subset. This approach ensures a balance between computational complexity and compression performance gains.
To further discuss this problem, we simplify the process following ~\cite{luo2024compressible}:
\begin{align}
    & \hat{y} = Q(y) =  Q(g_a(x)) \\ \notag
    & \hat{x} = g_s(\hat{y}) \\ \notag
    & \hat{z} = Q(z) = Q(h_a(x)) \\ \notag
    & p = h_s(\hat{z}) \\ \notag
\end{align},
where $g_a, g_s, h_a, h_s$ are the neural network parameters with encoder, decoder, the hyper-encoder and hyper-decoder at context model. $Q$ is the quantization. The internal feature are $y$, $z$ and $p$, and $x$ is the input image. 

\begin{align}
    & \mathcal{B} = AE_1(\hat{y}) + AE_2(\hat{z}) \\ \notag
    & AE_1 \sim [p, ctx(y,p) ]
\end{align},
$AE_{1,2}, \mathcal{B}$ are arithmetic coding and bitstream. The main bitstream is from $y$ with $AE_1$, so the most crucial aspect is modeling the data distribution to accurately predict the probability of $y$
from the aspect of data compression, achieving optimal compression rates. Hence, $ctx$ serves as the context model to analyze $p$ and $y$, ultimately providing the probabilities used by the $AE_1$ entropy coder. Thus $h_a, h_s, ctx$ are called hyper context. 

This paper separately evaluates the performance of increasing parameters for encoder, decoder and context model. We first discuss the ineffectiveness of scale for encoder, and then analyze its impact of parameter scales. However scales at decoder and context model is far from scaling law in large language model. 

\begin{figure}[t]
    \centering
    \includegraphics[width=0.495\linewidth]{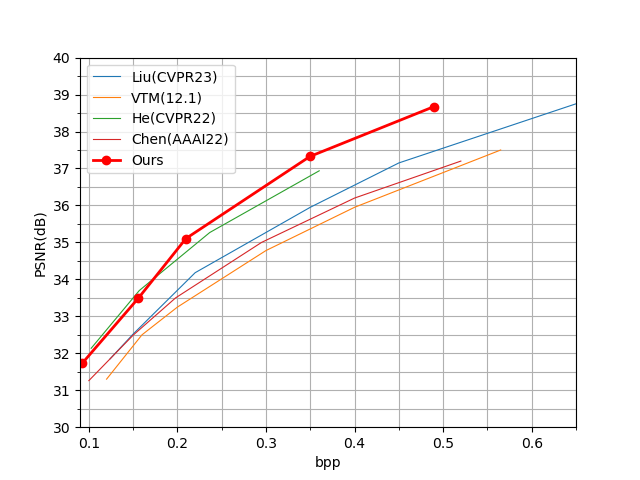}
        \includegraphics[width=0.495\linewidth]{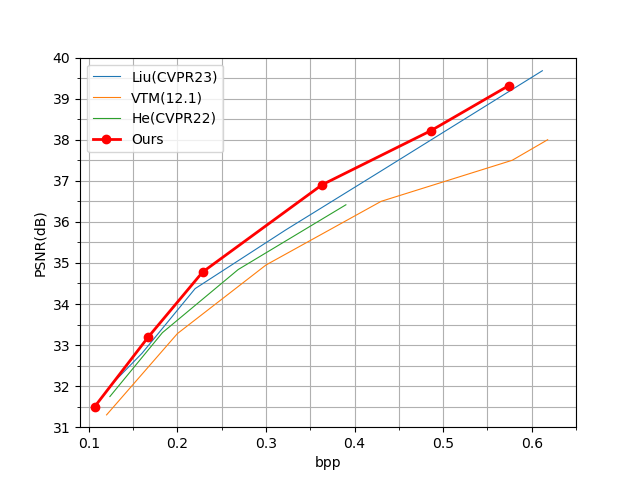}
    \caption{left:PSNR of CLIC Professional dataset. right:PSNR of Technick dataset.}
    \label{fig:clic_technick_psnr}
\end{figure}

\section{Motivation}
DCVC-FM~\cite{li2024neural} differs from encoding P-frame residuals in that it uses the I-frame as a context to probabilistically model the P-frame. As shown in the Tab.~\ref{tab:video}, the bit-rate of the P-frame is less than one-tenth of the I-frames. For the BQTerrace sequence, the P-frame achieves a higher PSNR than the I-frame at a bit rate of one two-thousandth, due to the relatively slow motion and similar background, allowing the entire I-frame to serve as a reference for the P-frame. This significantly reduces the bit rate. However, for other videos with intense motion, while the P-frame bit rate decreases, the PSNR also correspondingly decreases. On the other hand, if a single frame image has a bit rate below $0.001 bpp$, its PSNR drops well below 28 dB. These observations indicate that context is crucial for compression. To further compare, this paper uses ELIC to encode the remaining P-frames. It is evident that even with strong contextual representation, its performance is significantly inferior to the P-frame encoding in DCVC-FM. Specifically, in the BQTerrace sequence, DCVC-FM achieves $29dB$ at a bit rate of $0.0007bpp$, while ELIC requires a bit rate of $0.019bpp$ to achieve only $26dB$. Moreover, the remaining sequences in Tab.~\ref{tab:video} have the same trend. This further illustrates that \textbf{context is all you need}.

\section{Methods}

This paper adopts the structure with ELIC~\cite{elic}, with the modification of $g^{k}_{ch}$, where $k$ is the group of $y=g_a(x)$  split by the channel dimension. This paper introduces a plug-and-play module $ChannAttenBlock$ into $g^{k}_{ch}$ to enchance the context model as shown in Fig.~\ref{fig:channattenblock}. Besides this paper also combines adaptive exponential quantization following~\cite{luo2024super} with $y_{frac} = e^{(y_{frac} + a) *b } + c$, where $y_{frac}$ is the fractional part of $y$, and $step$ can control the extent of quantization, which can determine the number of constant $a, b, c$. the larger the step size, the higher the probability that the quantized value will be zero. And here is simple implementation Coding List~\ref{code:quantization}, a plugin can be plug-and-play for any LIC models, and $k$ is the index of channel group.   

\subsection{Scaling for Encoder}

\begin{align}
    & \mathcal{L} 
    = \mathcal{R} + \lambda \mathcal{D} \\ \notag
    & y^i = \sigma(w^i y^{i-1})
\end{align}

\begin{table}
\centering
\caption{The rate distortion with increase M of encoder and decoder}
\label{tab:encoder-decoder}
\begin{tabular}{c|c|c|c|c}
\hline \hline
M & 192 & 384 & 768 & 1536 \\ \hline \hline
bpp & 0.3881 & 0.3911 & 0.3834 & 0.3764 \\ \hline
PSNR & 33.3167 & 33.4058 & 33.4032 & 33.2826 \\ \hline
\end{tabular}
\end{table}

\begin{lstlisting}[language=Python, caption={The implementation of adaptive exponential quantization}, label={code:quantization},
                   basicstyle=\ttfamily\small, keywordstyle=\color{blue},
                   commentstyle=\color{gray}, stringstyle=\color{red}, numbers=left,
                   numberstyle=\tiny, stepnumber=1, numbersep=5pt, frame=single,
                   rulecolor=\color{black}]
#input: y_i, upper_bound, step, k
#output: y_o
def quantize(y_i, upper_bound=0.5, step=0.04, i=k):
    bias = upper_bound - step * i
    exp_half_b = (1. + np.sqrt(1 - 4*bias*(1-bias))) / (2*bias)
    exp_b = exp_half_b ** 2
    constant_b = np.log(exp_b)
    exp_ab = 1 / (exp_b - 1)
    ab = np.log(exp_ab)
    constant_a = ab / constant_b
    constant_c = - np.exp(constant_a * constant_b)
    y_frac = torch.frac(y_i)
    y_round = y_i - y_frac
    y_frac_flag = torch.where(y_frac>=0, 1., -1.)
    y_frac = torch.abs(y_frac)
    y_frac = torch.exp((y_frac + constant_a) * constant_b) + constant_c
    y_frac *= y_frac_flag
    y_o = y_round + y_frac
    return y_o
\end{lstlisting}

More experiments demonstrate that enlarging encoder makes a little difference to improve compression gain as shown in Tab.~\ref{tab:encoder-decoder}. Here is the training loss function for LIC , and we define the non-linear process in neural network:
$\mathcal{L}, \mathcal{R},\mathcal{D}$ are the total loss, the bit-rate and the reconstruction loss. $\lambda$ is the Lagrange multiplier to balance the bit-rate and reconstruction loss. To simplify the process of gradient propagation, we define the input of $i_{th}$ neural network layer with weight $w^i$ as $y^{i-1}$, which is the output of $i-1_{th}$ layer, besides $\sigma$ is the non-linear activation function. Thus we get the final output of whole neural network of LIC as $L$ is the last layer, and $y^o = y = g_a(x)$ which is the output of encoder, where we will restrict the bit-rate, here we neglect the affect of round operation for $y^o$. For decoder $g_s$ we have the following relationship:
\begin{align}
 & \frac{\partial \mathcal{L}}{ \partial w^i} = \lambda <\frac{\partial \mathcal{L}}{ \partial y^L}, y^{i-1} \frac{\partial y^L}{ \partial y^i} \mathcal{J}^i>  = \lambda <\frac{\partial \mathcal{D}}{ \partial y^L}, y^{i-1} \frac{\partial y^L}{ \partial y^i} \mathcal{J}^i>\\ \notag
 & \mathcal{J}^i = diag(\sigma^{'}(w_1^{i}y^{i-1}), \cdots, \sigma^{'}(w_d^{i}y^{i-1}))
\end{align}
For encoder $g_a$ we have the similar formulation:
\begin{align}
 \frac{\partial \mathcal{L}}{ \partial w^i} 
 & = \frac{\partial \mathcal{R}}{ \partial w^i} + \lambda <\frac{\partial \mathcal{D}}{ \partial y^L}, y^{i-1} \frac{\partial y^L}{ \partial y^i} \mathcal{J}^i>  \\ \notag
 & = <\frac{\partial \mathcal{R}}{ \partial y^o}, y^{i-1} \frac{\partial y^o}{ \partial y^i} \mathcal{J}^i> + \lambda <\frac{\partial \mathcal{D}}{ \partial y^L}, y^{i-1} \frac{\partial y^L}{ \partial y^i} \mathcal{J}^i> 
\end{align}

$\mathcal{R}$ does not influence the representation of weight in $g_s$ but affects the weight in $g_a$. Thus when we increase the parameter of encoder $g_{a}$ under the certain bit-rate, the information bottleneck may restrict the capability of $g_{a}$, especially for the low bit-rate, which only need less feature to reconstruct the whole image. To further uncover the mechanism we depict the target and result of entropy optimization. Here $y^o = g_a(x)$ is the symbol to be encoded, which is subject to Gaussian distribution:
\begin{align}
    & f\left(y^o\right)=\frac{1}{\sqrt{2\pi} \sigma}e^{-\frac{\left(y^o-\mu\right)^{2}}{2v^{2}}} \\ \notag
    & p(\hat{y^o}) \sim p(y^o) =\int_{\hat{y^o}-0.5}^{\hat{y^o}+0.5}f\left(y^o\right)d{y^o} \\
\end{align}
$\mu, \sigma$ is the mean and variance of symbol $y^o$, $p(*)$ is the probability function of maximum entropy distribution with constrains mean $\mu$ and variance $\sigma$.

\begin{align}
    & \mu = h_{s}(\hat{z})[0] \rightarrow y^o \\ \notag
    & \sigma = h_{s}(\hat{z})[1] \rightarrow 0 
\end{align}
$[0], [1]$ means $\mu, \sigma$ are share the input and trainable parameters, and divide the final output into two layers evenly, with the first layer being the mean and the second layer being the variance. When we optimize the object of bit-rate, we try to minimize bit-rate, $\mathcal{R} = -\log p(\hat{y^o}), \min  \mathcal{R} \rightarrow \max p(y^o)$, namely we meanwhile 
minimize the scale or the dynamic range of $y^o$, or minimize the value of variance $\sigma$, which means $y^o$ has less probability to obtain the value close to $\mu$. 

Thus quantization is an efficientive method to improce compression gain by choosing the important information to balance the bit-rate and distortion as shown in the right of  Fig.~\ref{fig:kodak-ablation}.

\begin{table}
\caption{The rate-distortion for scaling of context of GFLOPS~ with $1920 \times 1080$ resolution}
\centering
\label{tab:flops-id}
\scalebox{0.95}{
\begin{tabular}{c|c|c|c|c|c|cc}
\hline \hline
GFLOPS & channels & 128 & 360 & 620 & 1024 & 1280  \\ \hline \hline
\multirow{2}{*}{backbone} & $g_a$ &  933.78  & 943.81 & 962.30 & 1,006.24 &  1,043.67    \\  \cline{2-7}
 & $g_s$ &  933.78  & 943.81 & 962.30 & 1,006.24 & 1,043.67   \\ \hline
\multirow{3}{*}{hyper context} & $h_a$ & 8.23 & 14.72 & 22.00 & 33.31 &  40.48   \\ \cline{2-7}
 & $h_s$ & 16.68  & 35.09 & 55.71 & 87.75 &  108.06    \\ \cline{2-7}
 & $ctx$ & 140.98 & 198.23 & 262.39 & 362.08 & 425.25 \\ \hline \hline 
\multicolumn{2}{c|}{hyper context ratio(\%)} & 8.16 & 11.61 & 15.02 & 19.36 & 21.56  \\ \cline{2-7} \hline
& bpp & 0.5498 & 0.5504 & 0.5226 & 0.5131 & 0.5089  \\ \cline{2-7} 
& PSNR(dB) & 35.4269 & 35.5509 & 35.7491 & 35.4612 & 35.3188  \\ \hline \hline
\end{tabular}}
\end{table}

\subsection{Scaling for Decoder}

Decoder-only structure is to reconstruct or generate the input information. For large language model, generation is the key, and its range of output has a myriad of possibilities. Thus the scaling law can control the performance of large language model. However for LIC reconstruction has higher priority than generation. Different from in-context learning~\cite{xie2021explanation} at large language model, the varieties of output in LIC is constrained by $y^o$. But the reconstruction pixel is shift from the input pixel, reconstruction can be regarded as the restricted generation process. Thus the output space of LIC is also large but rather less than the  space of large language model, which can generate new objects or scenes. The number of parameter in decoder $g_s$ is vital for reconstruction or generation, but beyond of the discussion of this paper.  

\subsection{Scaling for Context}
Context model is the bridge between encoder and decoder. As discussed above, context model including $h_a, h_s$ is also entropy-constrained.  However increasing the parameters of context model facilitate the compression performance, especially for high bit-rate. Since the range of $y^o$ has become large, the value of $v$ or the output of context model has more probabilities. Thus the capability of $h_a, h_s$ must adopt more varieties. Then enlarging the context model is to accurately model the distribution of $y^o$.

As shown in Tab.~\ref{tab:flops-id} GFLOPS\footnote{PyTorch profiler which calculates the flops of PyTorch operators}, the performance of bpp-psnr improves as the proportion of context parameters increases. However, when the proportion of context parameters continues to increase, while the bit rate decreases, PSNR also decreases. Nevertheless, these points are located in the upper left corner of the rate-distortion curve, all belonging to the Pareto frontier.

\subsection{Scaling for Dataset}
To explore the compression limits on the Kodak dataset, we conducted direct overfitting on the Kodak data to investigate the boundary of PSNR. It was observed that at $0.1bpp$, the PSNR can reach up to $\sim 33dB$.
From the overfitting results of different structures, in  which this paper obtains $34.750dB$ at  $0.25bpp$ and ELIC obtains $35.526dB$ at $0.256bpp$, it can be seen that stronger structures lead to better performance, but this phenomenon is obvious in the medium and high bit rates, while the performance of the two is close in the low bit rate. It indicates that after the input information has been quantized and screened in the low bit rate, the importance of structure and parameters decreases, but in the medium and high bit rates, the importance of structure is highlighted, and the context is more important. The overfitting of the Kodak dataset enables all parameters to be approximated as context, which greatly improves the performance.

For CLIC2021 professional dataset, the overfitting models on Kodak get much worse rate-distortion performance with $27.06dB$ at $0.53bpp$, while the normal model with 8000 training dataset obtains $31.74dB$ at $0.09bpp$. 
This indicates that the generalization of training with a limited data set has great limitations, and that compression is to match patterns in images. Since 8000 training data contains almost all common image patterns, it can effectively compress most images. It can be further illustrated that LIC is data-driven and scene-adaptive, which also provides significant data support for the customized application of LIC.



Training with mixed dataset, this does not effectively enhance the performance. The context enhancement effect brought by the Kodak dataset is diluted by the other training sets, resulting in negligible bias of the model toward the Kodak dataset alone. Besides further scaling for dataset from $8000$ to $16000$ images sampled from Imagenet makes a little difference for representation since the information bottleneck as shown in the left of Fig.~\ref{fig:kodak-ablation}. Similar with traditional compression methods, the latents $y^o$ are orthogonal and sparse like discrete consine coefficients. The training dataset is treated as $\mathcal{X}$, and $\mathcal{Y}$ are the codebook extracted by $g_a$. This phenomenon also indicates that LIC attempts to learn the pattern relationships among all image pixels, with this relationship explicitly represented as context and structure in the context modeling, and implicitly represented within the model parameters.




\begin{figure}[t]
    \centering
    \includegraphics[width=0.75\linewidth]{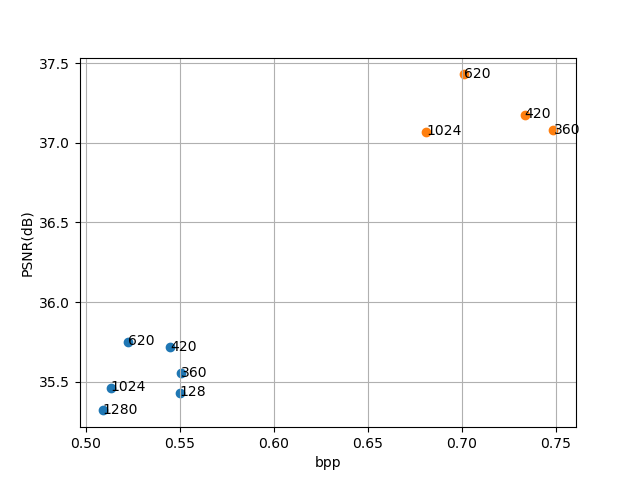}
    \caption{The rate-distortion performance with channel number of context model .}
    \label{fig:scalable}
\end{figure}

\section{Experiment}
\subsection{Datasets}
Training dataset is $8000$ images extracted from Imagenet~\cite{deng2009imagenet} with greater than $500\times 500$ resolution, and testing dataset is Kodak~\cite{kodak2022kodak}, CLIC 2021 professional dataset~\cite{theis2021clic} and Technick dataset with $1200 \times 1200$ resolutions. 

\subsection{Settings}
The training setting is the same as ELIC~\cite{elic} except for the number of channels $M=\{240, 320, 320, 360, 360\}$ for our standard model , with $\lambda=\{2, 4, 8, 20, 35\} \times 10^{-3}$ for lower bit-rate. Then $M=\{1024, 1024, 1024, 1024\}$ for our larger model , with $\lambda=\{8, 20, 35, 50\} \times 10^{-3}$ for higher bit-rate. While 'Overfitting' experiments are the same setting with the standard models with Kodak training data. Besides the quantization step default is $0.04$ for all $\lambda$. We train all models with $8000$ epochs, the first $6000$ epochs have learning rate with $1 \times 10^{-4}$, the remaining $2000$ epochs are $1 \times 10^{-5}$. 

\begin{figure}[t]
    \centering
\includegraphics[width=0.495\linewidth]{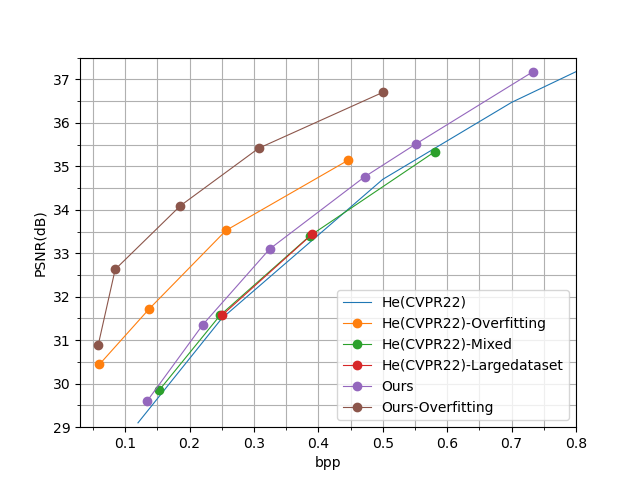}    \includegraphics[width=0.495\linewidth]{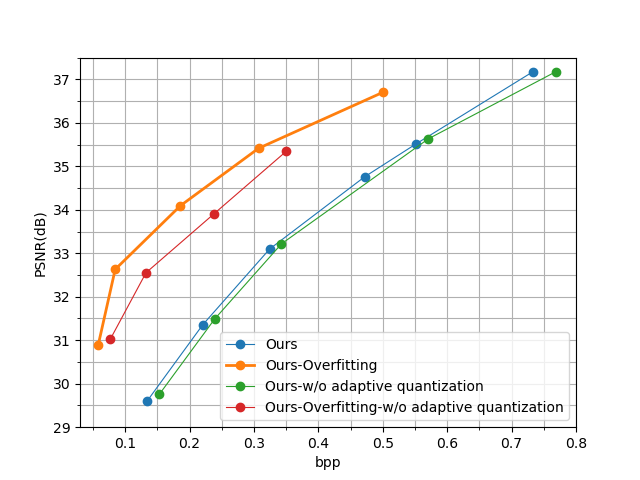}
    \caption{left:The influence of training dataset. right:The influence of adaptive quantization.}
    \label{fig:kodak-ablation}
\end{figure}

\subsection{Results}
As shown in Fig.~\ref{fig:kodak_psnr}, Fig.~\ref{fig:clic_technick_psnr}, this paper achieves the best bpp-PSNR over Kodak, CLIC2021 professional and Technick datasets. And we obtain $14.39\%$ BD-RATE gain and $69.22\%$ BD-PSNR gain over VVC, far higher than Liu~\cite{liu2023learned}. 

\subsection{Ablation experiments}

This paper discusses the training dataset, including overfitting, using larger-scale datasets, and training with a mix of the test and training sets. The results, as shown in the left of Fig. ~\ref{fig:kodak-ablation}, indicate that overfitting effectively enhances the context modeling capability. However, further enlarging the training dataset, even by mixing in the Kodak dataset with the training set, does not improve the compression performance on the Kodak dataset.

This paper also explores the impact of quantization as shown in the right of Fig.~\ref{fig:kodak-ablation}. The absence of quantization results in a decrease in PSNR, and in cases of overfitting, the PSNR decline is even more pronounced.

\section{Discussion}
Due to differences in network architecture and training methods, overfitting may yield varying results. This also demonstrates that this approach does not define the compression limits of LIC but rather proves that with more suitable context modeling is more efficient.

Although experiments involving simple scaling of the backbone demonstrate that further increasing the parameters does not continue to enhance compression gain in LIC, two issues need further research. First, CNN networks may not possess the scaling laws characteristics of transformers. Second, due to the smoothness and locality of images, the tokenization process differs from that of textual information and requires additional research. Besides the aforementioned entropy constraints, all these factors limit the further improvement of objective metrics of LIC.

{\small
\bibliographystyle{ieee_fullname}
\bibliography{neurips_2023}
}


\end{document}